# Identification and characterization of unique to human regulatory sequences in embryonic stem cells reveal associations with transposable elements, distal enhancers, non-coding RNA, and DNA methylation-driven mechanisms of genome editing

Gennadi V. Glinsky[1, 2, 3]

[1]The Stanford University School of Medicine,

Medical School Lab Surge Bldg, Room P214,

1201 Welch Road, Stanford, CA 94305-5494

Correspondence: gglinsky@stanfrod.edu

[2]Sanford-Burnham Medical Research Institute, 10901 North Torrey Pines Road

La Jolla, CA 92037

Email: gglinsky@sanfordburnham.org;

[3]Genlight Technology Corporation

939 Coast Blvd, Suite 4M

La Jolla, CA 92037

Phone: 858-401-3470

Email: genlighttec@gmail.com

www.genlighttechnology.com

www.facebook.com/genlighttech

**Running title:** Unique to human genomic regulatory sequences

**Summary**

Despite significant progress in structural and functional characterization of human genome, understanding of mechanisms underlying the genetic basis of human phenotypic uniqueness remains limited. We report that non-randomly distributed transposable element-derived sequences, most notably HERV-H/LTR7 and L1HS, are associated with creation of 99.8% unique to human transcription factor binding sites in genome of embryonic stem cells (ESC). 4,094 unique to human regulatory loci display selective and site-specific binding of critical regulators (NANOG, POU5F1, CTCF, Lamin B1) and are preferentially placed within the matrix of transcriptionally active DNA segments hyper-methylated in ESC. Unique to human NANOG-binding sites are enriched near the rapidly evolving in primates protein-coding genes regulating brain size, pluripotency lncRNAs, hESC enhancers, and 5-hydroxymethylcytosine-harboring regions immediately adjacent to binding sites. We propose a proximity placement model explaining how 33-47% excess of NANOG and POU5F1 proteins immobilized on a DNA scaffold may play a functional role at distal regulatory elements.

**Introduction**

Identification of genomic regulatory sequences contributing to development of species-specific phenotypic features during primate evolution remains a significant problem with important fundamental and translational implications. Molecular definition of mechanisms underlying the genetic basis of human phenotypic uniqueness represents particularly challenging problem. Despite remarkable progress in understanding the structural-functional diversity, transcriptional and regulatory complexities, and biological significance of non-protein-coding genomic regions, protein-coding genes remain the main focus of the experimental and theoretical analyses of the unique to human genetic traits. Accelerated evolution of a selected set of protein-coding genes involved in various aspects of nervous system development and biology has been demonstrated (Dorus et al., 2004; Zhang et al., 2011). Increased rates of protein evolution are most prominent for genes regulating brain size and behavior within the evolutionary lineage leading from ancestral primates to humans (Dorus et al., 2004). Recent experiments identify microRNA-associated regulation of brain gene expression as one type of human-specific regulatory mechanisms undergoing accelerated evolution (Somel et al., 2011).

Increasing contents and quality of genome and transcriptome sequence databases facilitated comparative genomics studies focused on identification of primate-specific genes in human genome. Taken together, these reports identified a few hundred candidate primate-specific genes for follow-up functional studies, demonstrated insertions of interspersed repeats in exons of 92% of primate-specific genes, documented placement of sequences derived from transposable elements (TEs) in 53% of primate orphan genes, and highlighted involvement of non-duplicated pericentromeric and subtelomeric regions of human genome in genesis of primate-specific genetic elements (Tay et al., 2009; Toll-Riera et al., 2012). However, systematic effort aiming at genome-wide identification and characterization of unique to human regulatory sequences such as transcription factor (TF) binding sites is lacking.

Repetitive and repeat-derived DNA sequences, including TEs, may account for up to two-thirds of the human genome (De Koning et al., 2011). TEs contribute to multitude of genomic regulatory functions and have been identified as a rich source for creation of new species-specific TF-binding sites in mammals (Wang et al., 2007; Bourque et al., 2008; Kunarso et al., 2010). Most recently, TEs have been implicated in the origin, regulation, and specification of long noncoding (lnc) RNAs (Kelley and Rinn, 2012; Kapusta et al., 2013). It has been documented that TEs are embedded within 83% of human lncRNAs and comprise 42% of lncRNA sequences, which is in sharp contrast to the small TE content within protein coding genes in human genome (Kelley and Rinn, 2012).

In human cells, retrotransposon activity is believed to be suppressed to restrict the potentially harmful effects of mutations on functional genome integrity and maintenance of genomic stability. Human embryonic stem cells (hESC) seem markedly different in this regard. Several studies demonstrated that hESC express mRNAs from both human-specific (L1HS) and older L1 sub-families of retrotransposons (Garcia-Perez et al., 2007; Macia et al., 2011) and activity of L1 retrotransposition is markedly elevated in hESC (Wissing et al., 2012). Endogenous retrovirus HERV-H RNA expression is markedly increased in hESC, and an enhanced rate of insertion of HERV-H sequences appears associated with binding sites for pluripotency transcription factors and lncRNAs (Santoni et al., 2012; Kelley and Rinn, 2012; Xie et al., 2013). These data suggest that TEs may contribute to creation of unique to human regulatory sequences associated with pluripotent stem cell phenotype. Activation of retrotransposons and lineage-specific repeat-driven dispersion of CTCF-binding sites has produced species-specific expansions of CTCF binding in genomes of rodents, dogs, and opossums (Schmidt et al., 2012). Importantly, this evolutionary engineering of new CTCF-binding events appears to create binding sites that function as chromatin domain insulators and transcriptional regulators (Schmidt et al., 2012). Surprisingly, there was no evidence found indicating the enrichment of CTCF-binding events within species-specific repeats in human or macaque (Schmidt et al., 2012).

In this contribution, we report identification and characterization of 4,094 unique to human regulatory loci that are preferentially placed within the matrix of DNA segments hyper-methylated in hESC and appear to operate in hESC in distinct and exclusive modes by engaging in selective and site-specific binding of critical regulatory proteins, NANOG, POU5F1 (OCT4), CTCF, and Lamin B1. Sequences of the unique to human TF-binding sites do not intersect any chains in the mouse, rat, chimp, gorilla, orangutan, gibbon, rhesus, and marmoset genomes. Our analysis reveals exclusive-to-primates DNA editing mechanisms involving RNA-associated DNA methylation at selected cytosine residues within LTR7 sequences coupled with methyl-cytosine deamination that are markedly overrepresented in human genome and appear involved in creation of regulatory sites with novel protein-binding specificities.

**Results**

**Repetitive elements are associated with more than 99% of unique to human TF-binding sites in hESC**

To identify unique to human regulatory sequences, we carried-out sequence homology profiling of 205,974 binding sites for NANOG-, OCT4-, and CTCF proteins detected in hESC (Kunarso et al., 2010) across human, rodent, and primate reference genome databases. In the hg18 human genome reference database, the LiftOver algorithm (http://genome.ucsc.edu/cgi-bin/hgLiftOver) identifies 29,130; 14,003; and 29,018 sequences of 200 bp in length centered at NANOG-, OCT4-, and CTCF binding sites that do not intersect any chains in the mouse and rat genomes (**Data Sets S1-S3**). Overall these sequences constitute 33% of all TF-binding events detected in hESC for NANOG and CTCF and 47% of all OCT4-binding sites (**Table 1**). Similar results were recorded when the analysis was performed employing the hg19 release of the human genome reference database, confirming the consistency of mapping and alignments for these regulatory sequences across databases. We defined these sequences as candidate primate-specific TF binding sites. To determine whether this collection of sequences contains a sub-set of unique to human regulatory loci, we searched for TF-binding sites that i) represent sequences uniquely mapped to a single genomic location in both hg18 and hg19 releases of the human reference genome database; and ii) do not intersect any chains in the rodent and primate genomes, including mouse, rat, chimp, gorilla, orangutan, gibbon, rhesus, and marmoset genomes. This analysis identifies 826; 2,386; and 591 unique to human binding events for NANOG, OCT4, and CTCF transcription factors, respectively (**Data Sets S4-S6**). Notably, these sets of unique to human regulatory sequences represent 0.9%, 8%, and 0.7% of all binding sites and 3%, 17%, and 2% of primate-specific binding sites for NANOG, OCT4, and CTCF transcription factors, respectively (**Table 1**).

To determine whether repetitive elements contributed to creation of unique to human TF-binding sites, we used the 200-bp window centered at the middle of the TF-binding sites and intersected these sequences with the RepeatMasker database track of the UCSC Genome Browser (http://www.repeatmasker.org/ ). Each overlapping event was tabulated and the numbers of overlaps of each transcription factor's binding events with specific repetitive elements were calculated. Notably, we observed that 99.8% (3,797 of 3,803) of unique to human TF-binding sites are embedded within repetitive elements, which is significantly more than expected by chance ($p \ll 0.0001$; hypergeometric distribution test). Our follow-up analyses indicate that the "99% rule" is not limited to the NANOG, OCT4, and CTCF binding sites and seems to have a broad relevance. All unique to human binding events identified for five different regulatory proteins (SOX2; RNAPII; TAF1; KLF4; and p300) map within repeat-derived sequences in the reference human genome database (**Table 1; Fig. 1**).

We found that 532, 116, and 104 of unique to human binding sites for NANOG, OCT4, and CTCF, respectively, are located within LINE and LTR sequences, suggesting that retrotransposons play a significant role in creation of unique to human regulatory sequences (**Data Sets S4-S6**). LINE- and LTR-embedded unique to human TF-binding sites constitute 64%, 5%, and 18% of all unique to human binding events for NANOG, OCT4, and CTCF, respectively. LINE and LTR elements appear to make equal contribution to creation of unique to human TF-binding sites for NANOG, OCT4, and CTCF (**Fig. S1**). Interestingly, a majority of LINE- and LTR-embedded unique to human TF-binding events in hESC is represented by transcriptionally active in human cells L1PA2 (196 events), L1HS (64), LTR7 (109), and LTR5_HS (81) retrotransposons (**Fig. S1**). Notably, unique to human TF-binding sequences are often embedded within full-length 6 Kb L1PA2 and L1HS sequences. For example, 95% (133 of 140) L1PA2-embedded NANOG-binding sites are located within the full-length L1PA2 sequences. Similarly, full length 6 Kb L1HS sequences with less than 1% divergence constitute 77% (34 of 44) of all L1HS sequences with unique to human TF-binding sites in hESC. Taken together with the recent observations that L1 retrotransposition is markedly elevated in hESC (Wissing et al., 2012) and mRNAs from both human-specific L1HS and older L1 sub-families of retrotrasposons are expressed in hESC (Garcia-Perez et al., 2007; Macia et al., 2011), these data support the idea that L1 retrotransposition is actively contributing to creation of unique to human regulatory loci.

**Unique to human TF-binding events manifest distinct regulatory patterns in hESC and differentiated cells**

One of the notable features of unique to human regulatory loci reported in this study is that many of the active in hESC TF-binding sites are not bound by TFs in differentiated cells. This observation is consistent with documented essential role of the NANOG and OCT4 in maintenance of the pluripotent state and the restriction of their functions to ESC. We observed a significantly diminished (p < 0.0001) CTCF-binding in differentiated human cells compared to hESC (**Fig. S1D**), suggesting that the function of a large number of CTCF-binging events is associated with the pluripotent state. Interestingly, loss of the CTCF occupancy in differentiated cells appears limited to the unique to human and shared with primates binding events (**Fig. S1D**), indicating that pluripotency-associated CTCF activity may be mediated by primate-specific binding sites. Recent experimental evidence supports the idea of specialized CTCF-dependent regulatory mechanisms that are critical for self-renewal and pluripotency of hESC. Balakrishnan et al. (2012) reported on functional analyses of individual CTCF binding sites in hESC by demonstrating specialized, site- and target gene-specific CTCF-dependent insulator, enhancer, or repressor activities.

NANOG functions in hESC seems particularly interesting in this context because nearly two-third (542 of 826; 64%) of unique to human NANOG binding events are embedded within LINE and LTR retrotransposons in contrast to only 116 (5%) and 104 (18%) binding sites for OCT4 and CTCF, respectively. These data suggest that there may be a selective pressure to retain NANOG-binding sites created by continuing activity of retrotransposons in hESC. Consistent with this line of arguments, 34 of 39 (87%) of TF-binding loci embedded within transcriptionally active full-length L1HS retrotransposons (**Data Set S4**) are occupied by NANOG, which is significantly more than expected by chance (p = 0.00086). In contrast, only seven (18%) full-length L1HS-embedded unique to human TF-binding sites are occupied by OCT4, five of which overlap with NANOG-binding sites. Consistently, only three (8%) full-length L1HS-embedded unique to human TF-binding sites are occupied by CTCF which is marginally less than expected by chance (p = 0.044). These conclusions remain statistically valid after accounting for five overlapping sites (p = 0.004; p = 0.27; p = 0.08; for NANOG, OCT4, and CTCF, respectively). We conclude that unique to human TF-binding loci appear to operate in hESC in distinct and exclusive modes by engaging in selective and site-specific binding of critical regulatory proteins, NANOG, OCT4, CTCF. Given the high diversity of TF-binding sites in mammalian genomes recognizing myriads of DNA elements and continuing expansion of TEs in human population, at least some of these correlations are likely to be chance events unrelated to natural selection. Further functional studies will be required to clarify the biological significance of the apparent structural differences between hESC and differentiated cells.

**Unique to human TF-binding sites exhibit distinct patterns of transcriptional activity and association with nuclear lamina**

Next we carried out a systematic survey of the genomic regulatory landscape around unique to human NANOG- and CTCF-binding sites to identify regulatory elements associated with unique to human TF-binding events. We observed the following genomic features characteristic of regions adjacent to unique to human TF-binding events: i) frequent location within lamin-associated domains (LAD; **Figs. 1 & S2**); ii) consistent transcriptional activity (**Fig. 1**); iii) apparent association with domains of differential DNA methylation in hESC (**Fig. 2**). An example of the enrichment of TF-binding sites within LAD is illustrated in **Fig. 1** for NANOG-binding sites embedded within full-length 6 Kb LINE transposons. LAD occupies 42.9% of human genome (Guelen et al., 2008). Therefore, based on the random distribution model the expected number of L1-embedded NANOG-binding sites located within LAD would be 72, which is significantly less than the observed number of 104 binding events located within LAD (p = 0.0007; **Fig. 1**). We observed a significant difference in representations of the NANOG-binding sites embedded within truncated L1 sequences and located either within or outside the LAD boundaries. Only 5% of all L1HS- and L1PA2-embedded NANOG-binding sites located within LAD are represented by truncated L1 sequences (**Fig. 1**). In contrast, 15% of L1 retrotransposon-associated NANOG-binding events are located within truncated L1 sequences outside of the LAD boundaries (p = 0.0386; **Fig. 1**). These data indicate that LADs represent genomic regions which favor the placement and/or retention of full-length L1 sequences harboring unique to human NANOG-binding sites. This phenomenon appears L1 retrotransposon-specific, because LTR7-embedded NANOG-binding events are distributed equally within and outside LADs, whereas LTR5_HS-associated NANOG-binding sites are preferentially located outside the LAD boundaries (p = 0.0338; **Fig. 1**). Similarly, the number of unique to human RNAPII-binding sites placed outside LADs is markedly enriched in both K562 and MCF cells and depleted within LADs in both cell types (**Fig. 1**), which is consistent with previous reports identifying LADs as chromosomal domains with the predominantly repressive chromatin environment and low transcriptional activity (Guelen et al., 2008; Peric-Hupkes et al., 2010).

Transcriptional activity is another notable feature of unique to human genomic regulatory loci embedded within LINEs and LTRs. Analysis of corresponding RNA sequences in the Expression tracks of the UCSC Genome Browser revealed that expression is not limited to established human cell lines such as GM12878, K562, and H1-hESC. Active transcription from selected genomic loci containing unique to human TF-binding sites embedded within LINE and LTR sequences is apparent in human brain, heart, breast, and lymph nodes. NANOG-binding sites embedded within LTR7 appear particularly active in H1-hESC (**Fig. 1**). RNA molecules are detected at nearly 40% of all LTR7-associated NANOG-binding events and LTR7-derived transcripts account for more than 75% of all LINE- and LTR-associated transcriptional events observed in H1-hESC (**Fig. 1**). Consistent with these observations, Lister et al. (2009) reported that in hESC a subset of differentially methylated domains (DMRs) co-localize with dense clusters of small RNAs that map to annotated Human Endogenous Retroviruses (HERVs). They found that in hESC the HERVs were less densely methylated and frequently associated with high downstream transcriptional activity, in contrast to the more methylated state and low proximal transcriptional activities in differentiated human cells (Lister et al., 2009). Alternatively, these distinct patterns of associations may be due to the differential retention of different classes of TEs in different genomic compartments. For example, the preferential retention of L1 insertion in LADs may be due to the fact that these regions consistently manifest reduced transcriptional activity that will likely to reduce potential deleterious effects caused by spurious TF binding. On the other hand, the fact that 40% of LTR7 insertions carrying NANOG-binding sites occur within transcriptionally-active regions might be due to a preferential insertion of such TEs in open chromatin. Follow-up mechanistic studies will be required to distinguish between these possibilities.

**Unique to human LADs and TF-binding sites are placed within the matrix of DNA methylation domains hyper-methylated in hESC**

Evidence of the frequent placement of the unique to human TF-binding sites within LAD suggests that at least some LAD in human genome may constitute a component of the unique to human genomic regulatory network. To test this hypothesis, we set out to look for unique to human LAD defined as human LAD sequences which are: i) conserved in mouse and human genomes; ii) located outside of the LAD boundaries in the mouse genome of the four distinct types of murine cells, namely mESC; Neural Precursor Cells, NPC; astrocytes; and Mouse Embryonic Fibroblasts, MEF (Guelen et al., 2008; Peric-Hupkes et al., 2010). Using these criteria, we identified 290 unique to human LAD representing evolutionary conserved genomic sequences that manifest the increased propensity of association to nuclear lamina in human cells (**Data Set S7**). Unique to human LADs constitute more than 21% of all LADs in human genome indicating that genome-wide interaction profiles with nuclear lamina and 3D folding patterns of human chromosomes are markedly distinct. The average length of unique to human LADs is 361 Kb which is 2.7-fold shorter than the average length of all LADs in human genome (980 Kb; p = 5.382E-62) and 25-40% of all LADs of nearly half human chromosomes (11 of 23; 48%) are represented by the unique to human LADs (**Figs. 2 & S3**). Unexpectedly, we observed a significant correlation between the chromosomal distributions of 290 unique to human LADs and either 29,018 primate-specific CTCF-binding sites (r = 0.632; p = 0.0012) or 29,130 primate-specific NANOG-binding sites (r = 0.640; p = 001). In contrast, no significant correlation was recorded between the chromosomal distributions of 290 unique to human LADs and 14,003 primate-specific OCT4-binding sites. These observations suggest that the observed correlations cannot be explained solely by the chromosome size effects and may indicate that mechanisms of creation and retention of novel regulatory elements in human genome are associated with common structural features. Survey of the genomic landscape in the vicinity of unique to human TF-binding sites reveals apparent co-localization with DNA methylation sites, suggesting that genomic regions of DNA methylation may be relevant to these processes.

We asked whether unique to human regulatory elements manifest patterns of consistent co-localization with domains of differential DNA methylation in hESC (Lister et al., 2009). We found that nearly 90% of unique to human LADs (257 of 290, 88.6%; p < 0.0001) intersect with continuous DNA segments termed Partially Methylated Domains (PMDs), which are hyper-methylated in hESC and hypo-methylated in differentiated human cells IMR90 (Lister et al., 2009). Similarly, 90.5% (535 of 591); 65.1% (538 of 826); and 94.6% (2258 of 2386) of unique to human binding events for CTCF, NANOG, and OCT4, respectively, are located within the PMDs (p < 0.0001 in all instances; **Fig. 2**). This analysis demonstrates that a vast majority of identified here unique to human regulatory elements are located within genomic regions which are hyper-methylated in hESC and hypo-methylated in differentiated human cells IMR90. PMDs are significantly enriched for genes that were more highly expressed in hESC compared to differentiated cells, indicating that the chromatin state within these regions in hESC is permissive for high transcriptional activity (Lister et al., 2009). In agreement with these findings, we found consistent evidence of active transcription from unique to human TF-binding sites in hESC which is particularly apparent for LTR7-embedded loci (**Fig. 1**).

A recent report demonstrates that PMDs cover 37% of genome in the full-term human placenta (Schroeder et al., 2013), indicating that function of these epigenetic regulatory domains is not limited to the cultured cells. We found that 482 (58.4%) of unique to human NANOG-binding sites are located within 264 placental PMDs documenting statistically significant association (p<0.0001) of unique to human TF-binding sites with PMDs identified in a normal human tissue. Notably, among seven different tissue-specific PMDs and HMDs (highly methylated domains) only neuronal-specific domains (N-HMDs) are enriched for genes located near unique to human NANOG-binding sites (2-fold enrichment; p=0.0094). These data are consistent with the hypothesis that unique to human regulatory loci are preferentially placed within differentially-methylated genomic regions, because N-HMD designates domains defined by sequencing of bisulfite-treated DNA (MethylC-seq) as PMD (hypo-methylated domains) in IMR90 fetal lung fibroblasts and placenta but identified as HMD (hyper-methylated domains) in SH-SY5Y neuroblastoma cells (Schroeder et al., 2013). Notably, genes located in N-HMDs had synaptic transmission and neuron differentiation functions and play a role in brain and embryo development (Schroeder et al., 2013), suggesting that expression of genes of these functional categories may be affected by close proximity of unique to human regulatory elements.

**Evolutionary conservation of LTR7 small non-coding RNA-encoding sequences in primates**

Our subsequent analytical effort was focused on 33 nt and 24 nt LTR7-derived sequences of small non-coding RNAs (LTR7 sncRNAs) which were often represented among LTR7-associated transcripts in hESC (**Figs. 1, 3 & S4**). We concentrated our initial analysis on the genome-wide assessment of the fully-conserved 33 nt and 24 nt LTR7 sncRNA-encoding loci defined as the full-length sequences with 100% identity, no gaps and no mismatches. We classified both sequences as primate-specific, because: i) there are no full-length 33 nt and 24 nt LTR7 sequences in either mouse or rat genomes; and ii) chromosomal positions for 89% of genomic loci encoding 33 nt LTR7 sncRNAs are conserved in primates' genomes with most significant similarities noted for the human genome with genomes of chimpanzee and gorilla (**Figs. 3 & S4**). Reflecting their common origin, chromosomal distributions of 33 nt and 24 nt sequences in human genome manifest highly correlated patterns (r = 0.865; p < 0.0001). However, there are 8.6-fold higher number of loci encoding 24 nt LTR7 sncRNAs than 33 nt transcripts in human genome (1,129 versus 131 loci, respectively, in hg19 human genome reference database), suggesting that there is strong selective pressure to retain 24 nt LTR7 sncRNA-encoding sequences. To assess this phenomenon further, the size distribution analyses in human and primate's genomes of fully-conserved LTR7-derived sncRNA-encoding loci of various lengths were carried out (**Figs. 3 & S4**). Remarkably, we observed essentially identical profiles of genome-wide size distributions of the LTR7 sncRNA-encoding loci in human, chimpanzee, and gibbon genomes (r = 0.998; p < 0.0001) with identical 24 nt and 18 nt sequences being observed as the predominantly conserved molecular entities in all instances (**Figs. 3 & S4**). The loci of 13 nt in length manifest seemingly random distributions of sequence identities without clear prevalence of a single molecular entity to exceed 20% of total genome-wide counts. Intriguingly, the number of conserved genomic loci encoding 24 nt or 18 nt LTR7 sncRNAs is ~2-20-fold higher in human genome compared to individual primate's genomes (**Figs. 3 & S4**), indicating that human genome is markedly enriched for genomic loci encoding LTR7-derived sncRNAs.

We explored further the apparent association of the expressed in hESC LTR7 sncRNAs (**Fig. 1**) with DNA methylation events by calculating the frequency of co-localization of sncRNA expression with methyl-cytosine in human cells (**Fig. S4**). We computed co-localization events using data visualization tools in the AnnoJ browser (http://neomorph.salk.edu/human_methylome ) which are designed based on a genome-wide single-base resolution human methylome map (Lister et al., 2009). We observed that 89% of loci display co-localization of DNA methylation events with sncRNA expression within examined 200 nt in length sequences (**Fig. S4**). Of note, co-localization of the sncRNA expression with non-canonical methylation sites at CHH or CGH sequences appears enriched at the statistically significant level (p = 0.0057; **Fig. S4**). Methylation at non-canonical mCH sites increases most rapidly during the primary phase of synaptogenesis in the developing postnatal human brain and a rapid mCH level increase occurs primarily in neurons (Lister et al., 2013).

**Non-coding RNA-associated methyl-C to T mutation mechanisms of genome editing in hESC**

The CpG sites are hyper-mutable because the C of CpGs is considered a preferred site of DNA methylation, and methyl-C (mC) is prone to mutate to T via spontaneous deamination (Razin and Riggs, 1980; Ehrlich and Wang, 1981). The net result is that CpGs are replaced over time by TpG/CpAs sequences and the overall mCpGs mutation rate is estimated at 10–50 times the rate of C in any other context (Razin and Riggs, 1980; Ehrlich and Wang, 1981; Coulondre et al., 1978; Duncan and Miller, 1980; Bulmer, 1986; Sved and Bird, 1990), or of any other base in the genome (Hwang and Green, 2004). Recently, Lister et al. (2009) reported that ~25% of all methylation events identified in hESC were in a non-CG context. Methylation in non-CG contexts seems specific to ESC because non-CG methylation disappeared upon induced differentiation of the hESC (Lister et al., 2009). These data suggest that the mutation-driving mechanism caused by the spontaneous deamination of mC to T may be relevant to the non-canonical methylation events occurring in hESC at non-CG (e.g., CHH and CHG, where H is any base) sequences. We noted that 24 nt LTR7 sncRNA sequence contains 11 potential non-CG methylation sites, indicating that a spontaneous mC deamination mechanism may be relevant for generation of the C to T mutations within 24 nt sequences. We tested this hypothesis by performing a systematic search for a single-site C to T mutant sequences of differing lengths in human genome (**Figs. 3 & S4**). Strikingly, these analyses identify in human genome 21,906 genomic loci encoding 12 to 24 nt in length single-site C to T mutants of the 24 nt LTR7 sncRNA-encoding sequence (**Fig. S4**). Notably, C to T mutations at different positions within the 24 nt LTR7 sequence are generated and/or retained in human genome with markedly different efficiency: mutants at C3, C4, C5, C9, and C10 sites are represented by 2,893-4,146 loci, whereas mutants at C1, C2, C7, C8, and C11 positions are represented by 426-998 loci. The single-site C to T mutants seem to create primate-specific, predominantly highly conserved sequences because chromosomal positions of more than 90% of examined mutant loci are conserved in genomes of human, chimpanzee, gorilla, orangutan, and gibbon (**Fig. S4**). Interestingly, the generation and/or retention of mutants harboring the second C to T mutation within the same 24 nt LTR7 sequence was negligible as we found relatively few loci containing double C to T mutants in human genome. Our analysis argues that mC to T mutations outside the canonical mCG sites may play an important role in continuing editing of TE-derived sequences in primates' genome. This mechanism of mutations appears linked in human genome to 25,756 DNA segments (3,850 genomic loci encoding wild-type LTR7 sncRNAs and 21,906 genomic loci encoding 12 to 24 nt in length single-site C to T mutants of the 24 nt LTR7 sncRNA) containing LTR7-derived 24 nt sequences (**Fig. S4**).

Of note, multiple sequence alignment analysis of the unique to human NANOG-binding sites embedded within the LTR7-derived sequences identifies motif logos which appear closely resembling previously reported consensus binding sequences of several transcription factors, including NANOG, POU5F1 (OCT4), GATA1, and FOXA (**Fig. S5**). UA9 motif logo (**Fig. S5,** panels **B & C**) was previously associated with TF-binding events of multiple regulatory proteins exclusively in H1-hESC, including NANOG, BCL11A, HDAC2, ESR1, GATA2, RXRA, TCF12, and a sub-set of HDAC2-associated POU5F1 binding sites (Wang et al., 2012). We noted that several nucleotide variations within the sequences resembling the diverse set of TF-binding site motif logos are consistent with C to T and G to A patterns of mutations associated with methyl-cytosine deamination, suggesting that these mechanisms may contribute to the creation of novel TF-binding sites (**Fig. S5**).

**Association of unique to human TF-binding sites with brain-specific protein coding genes with accelerated rates of evolution in primates**

An elite set of twenty-four genes regulating brain size and behavior in humans and manifesting markedly accelerated rates of protein evolution within the lineage leading from ancestral primates to humans has been identified (Dorus et al., 2004).This set of genes may play an important role in defining phenotypic uniqueness of Homo sapiens by contributing to a dramatic increase in size and complexity of human brain during evolution (Dorus et al., 2004). We were interested to determine how unique to human TF-binding sites are positioned in the genome relative to genomic coordinates of these genes. We consider the housekeeping genes as the appropriate control set of genes in this analysis because they perform the essential basic cellular functions conserved across different species and are likely to evolve under evolutionary constrains without the significant impact of positive selection. Consistent with these assumptions and in contrast to the nervous system-related genes, housekeeping genes have statistically undistinguishable evolutionary rates in primates and rodents (Dorus et al., 2004). Because the twenty-four gene set is characterized by high evolutionary rate and manifests marked evolutionary rate disparities between primates and rodents (Dorus et al., 2004), we further increase the stringency of the comparison by selecting a set of fifty-three genes with the highest rates of evolution among housekeeping genes in primates (**Figs. 4 & S6** ).

We calculated for each gene in both gene sets the number of unique to human TF-binding sites with genomics coordinates placed in a relative proximity to protein-coding gene boundaries defined in hg19 reference human genome database. We choose to define the quantitative limits of proximity by the metrics placing unique to human TF-binding sites closer to putative target genes than experimentally defined distances to nearest targets of 50% regulatory proteins analyzed in hESC (Guttman et al., 2011). For each gene of interest we identified all unique to human NANOG- and CTCF-binding sites having genomic distances between TF-binding sites and a putative target gene smaller than the mean value of distances to nearest target genes regulated by the protein-coding TFs in hESCs, which were experimentally determined by Guttman et al. (2011). We observed that housekeeping genes and brain-specific genes manifest similar association patterns with unique to human TF-binding sites measured by the number of binding events located within the boundaries of putative target gene-specific regulatory regions (**Table 2; Figs. 4 & S6**). On average, 4.00 and 3.98 unique to human TF-binding sites were located in close proximity to the rapidly evolving brain-specific and housekeeping genes, respectively. For genes in both brain-specific and housekeeping sets, the evolutionary processes appear to favor the retention and/or creation of unique to human NANOG-binding sites compared to CTCF-binding sites (**Figs. 4 & S6**). These data indicate that the overall patterns of placement and retention of unique to human TF-binding sites with respect to genomic coordinates of either brain-specific or housekeeping genes are similar. Intriguingly, this analysis reveals that brain-specific and housekeeping genes manifest remarkably different correlation profiles between the numbers of unique to human TF-binding sites and rates of protein evolution of individual genes (**Figs. 4 & S6**). In the brain-specific gene set, we observed a highly significant positive correlation ($r = 0.992$; $p = 0.0079$), that is genes with the high numbers of located in close proximity unique to human TF-binding sites manifest higher rates of protein evolution (**Figs. 4 & S6**). In striking contrast, in a housekeeping gene set, we find a highly significant negative correlation ($r = -0.999$; $p = 0.0007$), that is genes with the high numbers of located in close proximity unique to human TF-binding sites manifest lower rates of protein evolution (**Figs. 4 & S6**). Notably, we observed the most significant correlation in the brain-specific gene set between the evolutionary rates of proteins regulating brain size in humans and the number of unique to human NANOG-binding sites placed in close proximity to these genes (**Figs. 4 & S6**). These data suggest that there is a selective pressure to acquire and/or retain novel NANOG-binding sites located in close proximity to the rapidly evolving genes regulating the size of human brain. In contrast, novel NANOG-binding sites appear preferentially placed in close proximity to the housekeeping genes with low rates of protein evolution. Alternative possibility is that rapidly-evolving genes might be less constrained in humans and hence more tolerant to the transcriptional regulatory effect of nearby unique to human NANOG-binding sites, provided that this regulatory influence is important for neuronal gene expression. Therefore, this mechanism may be similar to the known depletion of TEs in promoters caused by the negative impact of TE insertions in genomic regulatory regions: TEs are preferentially found outside of promoter regions not due to selection for insertion in intergenic regions but to negative selection against insertion in promoters. It will be of interest to determine whether the association with rapidly evolving genes is a function of the new TF-binding sites or of TEs.

**Association of unique to human TF-binding sites with pluripotency lncRNAs and early developmental enhancers**

Many unique to human TF-binding sites manifest hESC-specific profiles of regulatory protein binding and transcriptional activities (**Fig. 1**). We found that disparities in transcriptional activities between hESC and differentiated cells are particularly evident for LTR7-derived sncRNAs (**Fig. 1**). In contrast to protein-coding genes, TEs are embedded within 83% of human lncRNAs and comprise 42% of lncRNA sequences in human genome (Kelley and Rinn, 2012), suggesting that TE activity is associated with the origin, regulation, and specification of lncRNA-encoding genes (Kelley and Rinn, 2012; Kapusta et al., 2013). Notably, a prominent direct correlation of the LTR7 (HERV-H) transcriptional regulatory signals with hESC-specific expression of lncRNAs has been reported (Kelley and Rinn, 2012). Taken together, these observations suggest that unique to human TF-binding sites, in particular, regulatory elements embedded within LTR7 sequences, may be relevant to the regulation of hESC-specific lncRNAs, namely pluripotency and lineage-specification lncRNAs (Guttman et al., 2011). Similarly to the analyses performed for protein-coding genes, we calculated for each lncRNAs the number of unique to human TF-binding sites with genomics coordinates placed in a relative proximity to lncRNA gene boundaries defined in hg19 reference human genome database. Remarkably, we observed that unique to human TF-binding sites are placed in close proximity to 15 of 25 (60%) of pluripotency lncRNAs (**Fig. S6**), which is significantly higher than expected by chance ($p \ll 0.0001$). Unique to human NANOG-binding sites comprise 88% (23 of 26) of all unique to human TF-binding events associated with pluripotency lncRNAs (**Table 2; Fig. S6**), whereas unique to human CTCF-binding sites are detected only in 3 of 26 instances (12%). Conducting similar analyses for 34 lineage-specification lncRNAs (Guttman et al., 2011), we observed the association of 14 lncRNAs (41%) with 27 unique to human TF-binding sites, 16 of which are NANOG-binding sites (**Table 2; Fig. S6**).

The spectrum of unique to human binding events associated with lineage-specification lncRNAs appears significantly different from that of pluripotency lncRNAs (**Table 2; Fig. S6**). Placement and/or retention of unique to human NANOG-binding sites appears enriched near pluripotency lncRNAs (p = 0.0059; **Table 2**). Unique to human NANOG-binding sites are found in close proximity to 15 of 15 (100%) pluripotency lncRNAs, whereas CTCF-binding sites are detected only in 4 of 15 (27%) instances (**Fig. S6**). In contrast, similar numbers of lineage-specification lncRNAs are co-localized with unique to human binding sites for either NANOG- or CTCF: the binding events are detected in 9 of 14 (64%) and 10 of 14 (71%) instances, respectively (**Fig. S6**). Sixteen unique to human NANOG-binding sites are detected near 14 lineage specification lncRNAs (59%; 16 of 27 TF-binding events), which is significantly less than expected by chance (p = 0.0029; **Table 2**). Unique to human CTCF-binding sites represented in 41% (11 of 27) of all events associated with lineage-specification lncRNAs (**Fig. S6**). Overall, we observed the mirror-image reciprocity of genomic placement and/or retention of unique to human TF-binding events between pluripotency and lineage specification lncRNAs (**Fig. S6**; p = 0.0276 for pluripotency-associated versus lineage specification-associated events).

Early developmental enhancers (EDEs) play important roles in human embryonic development and are distinguished in hESC by unique chromatin signatures and biological activities (Rada-Iglesias et al., 2011). Next we searched for unique to human regulatory elements associated with 7,596 EDEs (**Table 3**) previously identified in hESC (Rada-Iglesias et al., 2011). Our analysis identifies 1,594 primate-specific EDEs in human genome, which appears significantly smaller number of regulatory events compared to the numbers of primate-specific TF-binding sites (p < 0.0001 in all comparisons). In contrast to TF-binding sites, we found only seven unique to human EDEs, which represent 0.09% of all EDEs in human genome and appear markedly reduced compared to numbers of unique to human TF-binding events (p < 0.0001 in all comparisons; **Tables 1 & 3**). We identified forty-six EDEs spatially associated with fifty unique to human NANOG-binding sites reflecting statistically significant enrichment of regulatory loci co-localized within 10 Kb continuous genomic regions (**Table 3**).

**Association of unique to human NANOG-binding sites with functional distal hESC enhancers and 5-hydromethylcytosine**

Our analysis suggests that one of the mechanisms of bioactivity of unique to human NANOG-binding sites may be associated with functions of promoter-distal regulatory elements. To test this hypothesis, we assessed the co-localization pattern of unique to human NANOG-binding sites and distal regulatory elements utilizing a comprehensive database of hESC enhancers (Xie et al., 2013). We observed 431 co-localization events ($p=1.5667E^{-108}$) between 264 unique to human NANOG-binding sites ($p=2.99359E^{-22}$) and 331 hESC enhancers ($p=1.58558E^{-49}$) within 10 Kb continuous DNA segments in hESC genome. The placement enrichment metrics near hESC-restricted enhancers was significantly higher compared to all hESC enhancers: the enrichment values were higher 2.28-fold for unique to human NANOG-binding sites ($p<0.0001$), 1.69-fold for hESC enhancers ($p=0.0001$), and 1.71-fold for co-localization events ($p<0.0001$). Furthermore, we found that 39 unique to human NANOG-binding sites ($p=2.044E^{-13}$) are co-localized with hESC enhancers implicated in 122 high-confidence enhancer/protein coding gene regulatory pairs ($p=1.0924E^{-102}$).

Functional distal regulatory elements in hESC are markedly enriched for 5-hydroxymethylcytosine (5hmC), which is most enriched in regions immediately adjacent to sequence motifs of TF-binding sites (Yu et al., 2012). Using a base-resolution map of 5hmC in hESC genome, we determined that 5hmC is located near 46% (379 sites; $p=2.435E^{-53}$) and 21% (176 sites; $p=1.1613E^{-115}$) unique to human NANOG-binding sites within 1 Kb and 100 bp windows near unique to human NANOG-binding sites, respectively (**Figs. 5 & S7**). In contrast, we did not observe enrichment of 5hmC nucleotides near unique to human CTCF-binding sites within 1 Kb windows and found that only 15% (91 of 591) of CTCF-binding sites were located near 5hmC, which is significantly less than the number of unique to human NANOG-binding sites ($p<0.0001$). Similarly, we observed no significant enrichment of 5hmC nucleotides near unique to human POU5F1-binding sites and calculated that only 14.9% of POU5F1-binding sites were located within 1 Kb of 5hmC, which significantly less than the number of co-localization events documented for unique to human NANOG-binding sites ($p<0.0001$). These data indicate that a large number of unique to human NANOG-binding sites harbor 5hmC within the canonical CpG context immediately adjacent to TF-binding sequence motifs, because in hESC nearly all (99.89%) of 5hmCs exist in the canonical CpG context (Yu et al., 2012). Based on these observations, we propose that unique to human TF-binding sites are most likely to play a functional role in hESC at promoter-distal regulatory elements. We tested this hypothesis by building regulatory maps of selected genomic regions depicting all NANOG-binding sites, hESC enhancer/coding gene regulatory pairs, and 5hmC distributions near regulatory loci (**Figs. 5 & S7**). We observed that unique to human and primate-specific NANOG-binding sequences are located near hESC enhancers, whereas common with rodents NANOG-binding sites are placed next to TSS of protein-coding genes (**Figs. 5A & S7**).

Analysis of 5hmC sequences located near unique to human NANOG-binding sites revealed common patterns created by the frequent usage of particular sequences, examples of which are depicted by 5hmC symbols in **Figs. 5B, S7, S8**. A three-letter 5hmC symbol with one window in the center of a sequence due to asymmetrical placement of the middle 5hmC on the opposite DNA strand has been detected particularly often. 57% of sequences having at least three 5hmC nucleotides and harboring at least one 5hmC adjacent to unique to human NANOG-binding sites within 100 bp windows contain this symbol (**Figs. 5B, S7, S8**). We observed 81 instances of co-localization of at least one copy of this three-letter 5hmC symbol in sequences containing multiple 5hmC nucleotides within 1 Kb windows centered at 826 unique to human NANOG-binding sites (**Figs. S7 & S8**). This number is likely an underestimate because only high-abundance 5hmC are detectable at reported sequencing depth and significantly more sequencing is required for a high-confidence resolution of low-abundance 5hmC at a single-base precision (Yu et al., 2012). In contrast to unique to human NANOG-binding sequences, only eight 5hmC sequences within 1 Kb windows near 591 unique to human CTCF-binding sites contain this three-letter 5hmC symbol and NANOG protein is bound to four of these sites ($p<0.0001$). Similarly to CTCF-binding sites and in sharp contrast to unique to human NANOG-binding sequences, we observed only 20 events displaying this three-letter symbols of 5hmC sequences within 1 Kb windows near 2386 unique to human POU5F1-binding sites and NANOG protein is bound to 13 of these sites ($p<0.0001$). Taken together, these data suggest that identified here common patterns of 5hmC sequences may contain an information code for recruitment of NANOG protein to specific genomic loci.

**Discussion**

Phylogenetic conservation can be considered as a strong evidence for an important functional role of genetic regulatory elements. However, since species-specific regulatory events are by definition not conserved, it is difficult to differentiate the biological significance of such events from mere background noise until detailed functional experiments are performed. Multiple lines of indirect evidence support the concept of biologically important role of unique to human regulatory loci. Genome-wide proximity placement analyses of unique to human NANOG-binding sites revealed apparently non-random placement and significant enrichment near distinct genomic elements, including PMDs, N-HMDs, hESC enhancers, enhancers/coding gene regulatory pairs, asymmetrical 5hmC sites, pluripotency lncRNAs, EDEs, and rapidly evolving in primates protein-coding genes regulating brain size in humans. Consistent with the functional role of the unique to human NANOG-binding sites, multiple sequence alignment analysis identifies motif logos of the TE-embedded NANOG-binding sites which is closely resemble previously reported consensus binding sequences of the POU5F1 (OCT4) and NANOG transcription factors (**Fig. S5**). Furthermore, Ingenuity Pathway Analysis (IPA; http://www.ingenuity.com/) of coding genes that have within gene bodies or near gene boundaries unique to human NANOG-binding sites identifies a core set of 135 genes, which appear interconnected in multiple networks associated with the nervous system development and functions ($p=2.78E^{-03}$), embryonic development ($p=4.39E^{-04}$), behavior ($p=1.51E^{-03}$), and cardiovascular system development and functions ($p = 3.18E^{-03}$); (**Data Set S8 & Video**). One of the notable features of this set of genes is their apparent association with development of a broad spectrum of common human disorders, including cancer ($p=7.35E^{-08}$), cardiovascular diseases ($p=2.46E^{-04}$), reproductive system diseases ($p=2.16E^{-03}$), metabolic diseases ($p=2.56E^{-03}$); multiple neurological and psychological disorders, hereditary and developmental diseases, and many other disease states (**Data Set S8 & Video**).

    Identification of thousands primate-specific and unique to human sequences which appear to function as protein-specific binding elements in ESC raises the question regarding the potential functional impact of this large-scale chromatin re-wiring in human genome. One intriguing possibility would be the extension of time of the transcriptional control of pluripotency factors NANOG and POU5F1 at selected genomic loci during early

developmental stages. Dynamic local concentrations of the NANOG and POU5F1 proteins near chromatin regions enriched for TF-binding sites would be maintained at the biologically effective threshold levels for longer time periods during the gradual decline of the NANOG and POU5F1 expression in embryos. The proximity placement enrichment model does not require that all primate-specific and unique to human TF-binding sites function in a manner similar to the classic promoter, enhancer, repressor, or insulator elements. It indicates that 33-47% excess of NANOG and/or POU5F1 proteins (**Table 1**) that are immobilized on a DNA scaffold may affect the chromatin state and transcriptional regulatory landscape in selected genomic regions. For example, the large excess of TF-binding sites at selected genomic locations in primate genomes is likely to dramatically influence 3D chromatin conformation and affect transitions between distinct regulatory states as postulated in the strings and binders switch model of chromatin dynamics (Barbieri et al., 2012). Placement enrichment of primate-specific and unique to human TF-binding sites at selected genomic regions may facilitate chromatin remodeling and transition to the permissive chromatin states (Ernst and Kellis, 2013) enabling binding of multiple regulatory proteins and co-binding of many TFs at specific loci (**Figs. S9-S11**). Consistent with this model, we observed the statistically significant placement enrichment of unique to human NANOG-binding sites in close proximity to the genomic coordinates of rapidly evolving in primates protein-coding genes regulating brain size, pluripotency lncRNAs, hESC enhancers, and EDEs. Notably, many TF-binding enriched loci contain protein-coding genes that were implicated in regulation of critical elements of nervous system development such as quantity, density, and functionality of neurons, axons, synapses, and neurotransmitter receptors.

  Our analysis indicates that unique to human NANOG-binding sites are enriched within PMDs highly methylated in hESC and N-HMDs highly methylated in neuronal cells, suggesting that neuronal-specific highly methylated domains harboring synaptic transmission and neuron differentiation genes are established during the early embryonic stages of brain development. In agreement with this hypothesis, placement of unique to human NANOG-binding sites is enriched near 5hmC within the canonical CpG context immediately adjacent to TF-binding sequence motifs. Several lines of experimental evidence support this model. In human frontal cortex, 5hmC appears most abundant in neurons, rather than glial cells (Lister et al., 2013). Frontal cortex development is accompanied by increased enrichment of 5hmC at intragenic regions that are already hyper-hydroxymethylated at the fetal stage, demonstrating that adult patterns of genic 5hmC in frontal cortex are already evident in the immature fetal brain (Lister et al., 2013). In human brain, overall transcriptional activity is associated with intragenic 5hmC enrichment, with in utero establishment of adult 5hmC patterns for cell type–specific genes and loss of 5hmC enrichment associated with developmentally coupled transcriptional down-regulation (Lister et al., 2013).

According to the previous reports, TF-binding sites associated with repetitive elements constitute only relatively small fraction of all binding events. For example, repeat-associated binding sites accounted for 11.1% and 28.3% of all CTCF-binding events in genomes of human and mouse ESCs, respectively (Kunarso et al., 2010). Similarly, only 15% of 88,351 NANOG-binding sites and 22% of 29,740 OCT4-binding events in hESC are associated with repeats (**Table 1**). We observed a remarkably high association of unique to human TF-binding sites with repetitive elements: essentially all (99.8%; 3,797 of 3,803) unique to human regulatory sequences are embedded within repeats. Therefore, this contribution provides further support to the idea that TE represent a major evolutionary force for creation of new genomic loci with regulatory functions (Wang et al., 2007; Bourque et al., 2008; Kunarso et al., 2010) and highlights the potential mechanistic links between DNA methylation-associated genome editing and emergence of unique to human TF-binding sites.

Our analysis supports the hypothesis that genome editing mechanisms associated with active (autonomous) and passive retrotransposition, DNA methylation of TE-derived sequences, mC deamination and hydroxylation may facilitate the emergence of unique to human regulatory networks. Placement and/or retention of unique to human NANOG-binding sites within regulatory networks seems to favor close proximity to the genomic coordinates of rapidly evolving in primates protein-coding genes regulating brain size, pluripotency lncRNAs, hESC enhancers, and EDEs (**Tables 2 & 3; Figs. 4, 5 & S6**). Our analysis suggests that genomic regions which are hyper-methylated in hESC (PMD domains) may provide a permissive chromatin environment for unique to human regulatory sites. PMD domains are significantly enriched for genes that are over-expressed in hESC compared to differentiated cells, indicating that the chromatin state within these regions is permissive for high transcriptional activity in hESC (Lister et al., 2009). Collectively, these data support the hypothesis that unique to human TF-binding sites function at distal regulatory elements within networks regulating nervous system development during embryogenesis. Targeted gain- and loss-of-function follow-up experiments are required to determine whether these unique to human genomic regulatory networks exist and operate. In this context, experimental analyses of potential functional connectivity between unique to human TF-binding sites and brain-specific protein-coding genes with markedly accelerated evolution in the primate ancestor to human lineage are of particular interest.

This contribution provides possible answers to important questions why and how TE-derived sequences become functional DNA in the human genome. Several distinct mechanisms of mutations within repeat-derived sequences may facilitate the LINE and LTR transposon-driven creation of the NANOG, OCT4, and CTCF binding sites at new genomic locations. The intrinsic properties of reverse transcriptase to generate a high error rate when transcribing RNA into DNA is likely to represent one of the major source of mutations since, unlike any other DNA polymerases, reverse transcriptase has no proofreading ability. The intrinsically high error rate of reverse transcription associated with active and passive retrotransposition coupled with DNA methylation, mC deamination and hydroxylation would facilitate accumulation of mutations at an accelerated rate, thus providing a molecular basis for emergence, disappearance, and selection of novel TF-binding sites. Continuing cycles of LTR7 activity seems to operate within this mechanistic context as exquisite, primate-specific, selection-enabled designer and eraser of regulatory elements at new genomic locations.

CpG methylation and deamination plays a crucial role in the inactivation of transposons and protecting mammalian genomes from their harmful mutational activity (Yoder et al., 1997). Methylation and deamination of CpGs embedded within Alu transposons in the human genome resulted in generation of thousands of p53-binding sites with the preferred core motif composed of CpA and TpG dinucleotides (Zemojtel et al, 2009). It has been demonstrated that CpG deamination events may create TF binding sites with much higher efficiency than other single nucleotide mutational events (Zemojtel et al, 2009). Evolutionary analysis of TF binding sites in ESC is consistent with the idea that CpG deamination is a major contributor to creation of novel binding sites for NANOG, OCT4, and CTCF (Kunarso et al., 2009; Zemojtel et al., 2011). These naturally-occurring genome editing mechanisms may play an important selection-supported biological role in human evolution by markedly increasing a combinatorial regulatory complexity of individual genomes and enhancing phenotypic diversity of individual cells within populations. Transient activation of these mechanisms in embryogenesis may contribute to development of individually unique profiles of performance, fitness, longevity, and adaptation potentials of multicellular organisms during the ontogeny.

**Experimental Procedures**

We utilized solely publicly available data sets and resources as well as methodological approaches and a computational pipeline validated for primate-specific gene discovery (Tay et al., 2009; Kent, 2002; Schwartz et al., 2003). The analysis is based on the UCSC LiftOver conversion of the coordinates of human blocks to non-human genomes using chain files of pre-computed whole-genome BLASTZ alignments with a minMatch = 0.95 and other search parameters in a default setting (http://genome.ucsc.edu/cgi-bin/hgLiftOver). Extraction of BLASTZ alignments by LiftOver algorithm for a human query generates a LiftOver output "Deleted in new" which indicates that human sequence does not intersected with any chains in a given non-human genome. This indicates the absence of the query sequence in the subject genome and was used to infer the presence or absence of the human sequence in the non-human genome. Unique to human regulatory sequences were manually curated to validate their identities and genomic features using BLAST algorithm and the latest releases of the corresponding reference genome databases as of April-May, 2013.

Data sets of NANOG, POU5F1, and CTCF binding sites in hESC were reported previously (Kunarso et al., 2009) and are publicly available. RNA-Seq datasets were retrieved from the UCSC data repository site (http://genome.ucsc.edu/; Meyer et al., 2013) for visualization and analysis of cell type-specific transcriptional activity of defined genomic regions. Genome-wide map of human methylome at a single-base resolution was reported previously (Lister et al., 2012) and is publicly available (http://neomorph.salk.edu/human_methylome). The histone modification and transcription factors ChIP-Seq data sets for visualization and analysis were obtained from the UCSC data repository site (http://genome.ucsc.edu/; Rosenbloom et al., 2013). Genomic coordinates of the RNAPII-binding sites determined by the ChIA-PET method were obtained from the saturated libraries data sets for MCF7 and K562 human cells (Li et al., 2012). Genome-wide maps of interactions with nuclear lamina defining genomic coordinates of human and mouse LADs were obtained from previously published and publicly available sources (Guelen et al., 2008; Peric-Hupkes et al., 2010). We estimated the density of TF binding to a given segment of chromosomes by quantifying the number of protein-specific binding events per 1 Mb and 1 Kb consecutive segments of selected human chromosomes and plotting the resulting binding sites density distributions for visualization. Visualization of multiple sequence alignments was performed using WebLogo algorithm (http://www.weblogo.berkeley.edu/logo.cgi ). Consensus TF-binding sites motif logos were previously reported (Kunarso et al., 2010; Wang et al., 2012; Ernst and Kellis, 2013).

**Statistical analyses of the publicly available data sets**

All statistical analyses of the publicly available genomic data sets, including error rate estimates, background and technical noise measurements and filtering, feature peak calling, feature selection, assignments of genomic coordinates to the corresponding builds of the reference human genome, and data visualization were performed exactly as reported in the original publications and associated references linked to the corresponding data visualization tracks (http://genome.ucsc.edu/). Any modifications or new elements of statistical analyses are described in the corresponding sections of the paper. Statistical significance of the Pearson correlation coefficients was determined using the GraphPad Prism version 6.00 software. We calculated the significance of the differences in the numbers of events between the groups using two-sided Fisher's exact test and the significance of the overlap between the events using the hypergeometric distribution test (Tavazoie et al., 1999).

**Supplemental Information**

Supplemental information includes Supplemental Data Sets S1-S8; Supplemental Figures S1-S11; Supplemental Video and can be found with this article online at Cell Reports website.

**Author Contributions**

This is a single author contribution. All elements of this work, including conception of ideas, formulation and development of concepts, execution of experiments, analysis of data, and writing the paper, were performed by the author.

**Acknowledgements**

This work was made possible by the open public access policies of major grant funding agencies and international genomic databases and willingness to share the primary research data by many investigators worldwide. I would like to thank you many anonymous colleagues for their valuable critical contributions during the peer review process of this work.

**Figure legends**

**Figure 1.** Characterization of genomic features associated with unique to human NANOG- and CTCF-binding sites.

Location of full-length L1 TE sequences containing unique to human NANOG-binding sites is enriched within lamina-associated domains, LADs (**A**). In hESC genome, there are 184 unique to human NANOG-binding sites that are embedded within 167 full-length and 17 truncated L1HS & L1PA2 sequences, 110 of which are located within LAD (**A**, top panel). 104 of 110 (95%) of L1HS & L1PA2 sequences containing unique to human NANOG-binding sites and located within LADs are preserved as the full-length (5,962-6,189 bp) L1 retrotransposons (**A,** top panel), indicating that the conservation of the full-length L1 TE containing unique to human NANOG-binding sites seems significantly higher within LADs. In contrast, a majority of LTR5_HS-embedded unique to human NANOG-binding sites is placed outside LADs, while LTR7-embedded unique to human NANOG-binding sites are equally distributed within and outside LADs (**A,** bottom panel).

Distinct placement patterns within LADs of 446 unique to human TF-binding sites for five different regulatory proteins (**B**). Note that significantly higher fraction than expected by chance of RNAPII-binding sites is located outside of LADs (**B**), in contrast to other TF-binding sites.

LTR7-derived small non-coding RNAs (sncRNAs) represent the predominant species of sncRNAs generated by transcriptional activity of the unique to human NANOG-binding sites in hESC and transcribed from the loci equally distributed within and outside LADs (**C, D**).

**Figure 2.** Unique to human regulatory elements are associated with genomic regions hyper-methylated in hESC.

Unique to human LADs, comprising 21% of all nuclear lamina-associated genomic regions in human genome, are represented on all human chromosomes, constitute more than a quarter of LADs on 50% of human autosomes, are significantly smaller in size compared to all LADs in human genome (**A**), and manifest highly correlated patterns of chromosomal distributions with 29,018 CTCF-bound and 29,130 NANOG-bound primate-specific TF-binding sites (**Figure S3**).

Genomic coordinates of 4,094 unique to human regulatory loci representing four distinct classes of genomic regulatory elements are enriched within chromosomal regions hypermethylated in hESC compared to differentiated human cells and designated Partially Methylated Domains, PMDs (**B-E**).

**Figure 3.** Analysis of LTR7-derived sncRNAs transcribed from unique to human regulatory loci reveals evolutionary conservations of wild-type sequences.

Distinct molecular variants of the 33 nt and 24 nt LTR7 sncRNAs manifest primate-specific evolutionary spectrum of chromosomal distributions (**A-D**) and highly correlated patterns of chromosomal distributions in human genome (r = 0.865; p < 0.0001; **Figure S4**). Size distribution analyses in human and primate's genomes of fully-conserved LTR7 sncRNA-encoding loci of various lengths reveal essentially identical profiles of genome-wide size distributions of the LTR7 sncRNA-encoding loci in human, chimpanzee, and gibbon genomes (**A**) with identical 24 nt and 18 nt sequences representing the predominantly conserved molecular entities (**A, B**), which are significantly over-represented in the human genome compared to primates' and rodents' genomes (**A, B, D**). Panels **A** and **D** show results of size distribution analysis (**A**) and linear regression analysis (**D**) in rodent, human and primate's genomes of fully-conserved LTR7 sncRNA-encoding loci of differing lengths. Note essentially identical profiles of genome-wide size distributions of the LTR7 sncRNA-encoding loci in human, chimpanzee, and gibbon genomes (**C**, bottom panel; r = 0.998; p < 0.0001) with identical 24 nt and 18 nt sequences representing the predominantly conserved molecular entities (**A, D**), which are significantly over-represented in the human genome compared to primates' and rodents' genomes. Panel **C** (top two panels) shows evolutionary conservation in primates of chromosomal positions of the wild-type 33 nt LTR7 sncRNA-encoding sequences.

**Figure 4.** Association of unique to human NANOG-binding sites with rapidly evolving protein-coding genes regulating brain size in humans.

24 brain-specific and 53 housekeeping rapidly-evolving in primates protein-coding genes manifest distinct patterns of association with unique to human NANOG-binding sites (**A, C, D**). While placement of unique to human NANOG-binding sites is similarly enriched in close proximity to both brain-specific and housekeeping protein-coding genes (**A,** top panel; **Figure S6**), the placement of unique to human NANOG-binding sites is enriched near brain-specific genes having highest Ka/Ks ratios (**C**) defining most rapidly evolving genes and housekeeping genes with the lowest Ka/Ks ratios (**D**) reflecting their smallest rates of protein evolution (**A,** bottom panel; **C, D**). Note the significant positive correlations between the primates' Ka/Ks values and proximity placement of unique to human NANOG-binding sites for rapidly evolving protein-coding genes regulating brain size in humans (**B**).

**Figure 5.** hESC genomic maps of NANOG-binding sites, adjacent 5hmC, and functional hESC enhancer/gene pairs involved in enhancer-coding gene interactions.

Snapshots of base-resolution 5hmC maps in H1-hESC genome in close proximity to the unique to human NANOG-binding sites located near functional hESC enhancers and involved in high-confidence enhancer-target gene interactions as reported by Xie et al. (2012). Only predicted enhancer-gene pairs with p-value <=0.0001 were considered in the analysis. Also shown are genomic positions of the unique to human NANOG-binding sites (horizontal red bars), hESC enhancers, transcription start sites (TSS) of enhancer-targeted protein-coding genes, and maps of hESC enhancer-gene interaction pairs in H1-hESC (semi-transparent green arks). Positive blue bar values indicate 5hmC contents on the (+) Watson strand of DNA, whereas negative values indicate 5hmC contents on the (-) Crick strand. For 5hmC values, the vertical axis limits are -50% to +50%. Note that unique to human NANOG-binding sites (red horizontal bars marked by the NANOG signs) are placed in the heart of the large-scale hESC regulatory domains near functional distal enhancers within high-complexity multi-dimensional regulatory networks of interacting enhancer-gene pairs (**A**) as well as within bi-directional (**Figure S7**) distal regulatory domains. Examples of 5hmC patterns identified near unique to human NANOG-binding sites on all human chromosomes are shown in **B** and **Figure S7**. The exact genomic positions of 5hmC are indicated by blue vertical bars, NANOG-binding sites coordinates are depicted by red horizontal bars (unique to human loci), red arrows (primate-specific sequences), and red stars (common with rodents sites) based on hg18 release of the human genome reference database. Note that unique to human and primate-specific NANOG-binding sequences are located near hESC enhancers, whereas common with rodents NANOG-binding sites are placed next to TSS of protein-coding genes. Examples of common 5hmC three to eight letter symbols utilized to create a large variety of 5hmC patterns near unique to human NANOG-binding sites are shown in the yellow box (**B**, top left panel).

**Table 1.** Summary of the search for unique to human TF-binding sites.

| Regulatory proteins | hESC genome | Associated with repeats | Percent within repeats | Primate-specific sites | Unique to human sites | Unique to human sites within repeats | Percent within repeats | LTR/LINE embedded | Percent within LTR/LINE |
|---|---|---|---|---|---|---|---|---|---|
| Nanog | 88351 | 13200 | 14.9 | 29130 (33%) | 826 (0.9%) | 824 | 99.8 | 532 | 64.4 |
| CTCF | 87883 | 10021 | 11.4 | 29018 (33%) | 591 (0.7%) | 590 | 99.8 | 104 | 17.6 |
| POU5F1 | 29740 | 6619 | 22.3 | 14003 (47%) | 2386 (8%) | 2383 | 99.9 | 116 | 4.9 |
| RNAPII | 30585 | ND** | ND** | 12012 (39%) | 319 (1%) | 319 | 100.0 | 5 | 1.6 |

*Nanog, CTCF, & POU5F1 binding sites were determined by ChIP-Seq method in H1-hESC line (Kunarso et al., 2010; hg18 database counts); RNAPII-binding sites were determined by ChIA-PET method in MCF7 & K562 cell lines (Li et al., 2012; hg19 database counts); **ND, not determined.

**Table 2.** Association of unique to human NANOG-binding sites with rapidly evolving in primates brain-specific and housekeeping protein-coding genes and pluripotency long non-coding RNAs.

| Genes | Number of genes in the genome | Number (percent) of genes associated with unique to human Nanog binding sites | Number of associated unique to human TF binding sites | Binding site per gene ratio | Number (percent) of associated unique to human Nanog binding sites | P value* |
|---|---|---|---|---|---|---|
| Brain-specific genes | 24 | 20 (83%) | 115 | 5.8 | 86 (75%) | <<0.0001 |
| Housekeeping genes | 53 | 46 (87%) | 214 | 4.7 | 176 (82%) | <<0.0001 |
| Pluripotency lncRNAs | 25 | 15 (60%) | 26 | 1.7 | 23 (89%) | 0.0059 |
| Lineage-specification lncRNAs | 34 | 14 (41%) | 27 | 1.9 | 16 (59%) | 0.0029** |

* Hypergeometric distribution test; **Less than expected by chance; <<0.0001 designates p = 0 at a decimal place 30.

**Table 3.** Association of unique to human NANOG-binding sites with early developmental enhancers (EDEs) in hESC.

| EDE classification | Number of EDEs in the genome*** | Number (percent) of primate-specific EDEs | Number of unique to human EDEs | Number* of EDEs near to unique to human Nanog binding sites | P value** | Number of EDEs-associated unique to human Nanog binding sites | P value** |
|---|---|---|---|---|---|---|---|
| Class I | 5116 | 1157 (22.6%) | 6 | 31 | 1.89E-06 | 36 | 9.21E-09 |
| Class II | 2285 | 421 (18.4%) | 1 | 14 | 0.000873 | 13 | 0.002296 |
| Class II to I | 195 | 16 (8.2%) | 0 | 1 | 0.291112 | 1 | 0.291112 |
| Total | 7596 | 1594 (21%) | 7 | 46 | 8.15E-09 | 50 | 1.29E-10 |

* Number of enhancers and binding sites with overlapping genomic coordinates within continuous 10 Kb regions;
** Hypergeometric distribution test; *** Genomic coordinates and classification of the early developmental enhancers were reported in Rada-Iglesias et al. (2011).

**Inventory of Supplemental Information.**

Supplementary Data are available upon request:

Supplemental Data Sets S1-S8 (Excel Files);

Data Set 1: Primate-specific Nanog-binding sites

Data Set 2: Primate-specific Oct4-binding sites

Data Set 3: Primate-specific CTCF-binding sites

Data Set 4: 826 unique to human Nanog-binding sites

Data Set 5: 2386 unique to human Oct4-binding sites

Data Set 6: 591 unique to human CTCF-binding sites

Data Set 7: 290 unique to human LADs

Data Set 8: 198 genes identified by the Ingenuity pathway analysis

Supplemental Figures S1-S11;

Supplemental Video:

Ingenuity Pathway Analysis identifies network of 135 genes associated with unique to human NANOG-binding sites

**Supplemental Figure Legends**

**Supplemental Figure S1.** Related to Table 1.

Identification and characterization of 3,803 unique to human NANOG, OCT4, and CTCF binding sites in human embryonic stem cells.

Multi-species sequence homology profiling of 205,974 sequences centered at NANOG-, OCT4-, and CTCF-binding sites in hESC identifies 29,130; 14,003; and 29,018 primate-specific (**A**) and 826; 2,386; and 591 unique to human binding events for NANOG, OCT4, and CTCF transcription factors (**B**). 532, 116, and 104 of unique to human binding sites for NANOG, OCT4, and CTCF, respectively, are located within LINE and LTR sequences (**C**), a majority of which is represented by transcriptionally active in human cells L1PA2 (196 events), L1HS (64), LTR7 (109), and LTR5_HS (81) retrotransposons (**C**). Panel (**D**) shows markedly diminished numbers of active unique to human CTCF-binding sites embedded within LINE and LTR retrotransposons in differentiated human cells compared to the hESC. Note highly correlative patterns of chromosomal distributions of primate-specific CTCF-binding sites in differentiated human cells compared with the hESC (r = 0.946; p < 0.0001), indicating that placement patterns of primate-specific TF-binding sites in human genome reflect chromosome size.

**Supplemental Figure S2.** Related to Figure 1.

Chromosomal distributions of 1,417 unique to human NANOG (826 events) and CTCF (591 events) binding sites display highly correlated patterns with lamin-associate domains (LADs) on human chromosomes (left top panel).

LAD index values were calculated for each chromosome as a sum of the percentage of a given chromosome length within LADs and cumulative size of DNA sequence of a given chromosome within LADs (in Mbp). For comparison, the correlation between the chromosomal distributions of TF-binding sites and chromosome size is only marginally significant (right top panel; r=0.44; p=0.034). Chromosomal distributions of 29,130 primate specific NANOG-binding sites manifest highly correlated patterns with both LAD index values (bottom left panel; r=0.854; p<0.0001) and size of human chromosomes (right bottom panel; r=0.923; p<0.0001).

**Supplemental Figure S3.** Related to Figure 2.

Chromosomal distribution analyses of 290 unique to human LADs and 945 common with mouse LADs. Unique to human LADs comprise 21% of all nuclear lamina-associated genomic regions in human genome (**A**); they are represented on all human chromosomes, constitute more than a quarter of LADs on 50% of human autosomes (**A, B**), are significantly smaller in size compare to all LADs in human genome (inset in **A**), and manifest highly correlated patterns of chromosomal distributions with primate-specific CTCF (29,018 sites, panel **C**) and NANOG (29,130 sites, panel **D**) TF-binding sites, respectively.

**Supplemental Figure S4.** Related to Figures 2 & 3.

Characterization of wild-type and single-site mC to T mutants of LTR7 sncRNA-encoding loci revealed non-coding RNA-associated DNA methylation mechanisms of genome editing in hESC.

Analysis of genomic sequences encoding single-site mC to T mutants of the 24 nt LTR7-encoding sncRNAs reveals evolutionary conservations of mutant sequences (**A-D**). Mapping of eleven mC to T mutation sites within 24 nt LTR7 sncRNA-encoding sequences (**A**) manifesting statistically distinct patterns of retention and/or creation in human genome of the single-site mC to T mutants of differing lengths within the reverse (C1-C8) and forward (C9-C11) strands 24 nt LTR7 sequence (**B**). Conservation of primate-specific evolutionary spectrum of chromosomal distributions of the single-site mC to T mutants of the 24 nt LTR7 sncRNA-encoding sequence (**D**). Panel **C** shows the results of linear regression analysis of genome-wide size distributions of the 3,850 wild-type and 21,906 single-site mC to T mutants of the 24 nt LTR7 loci and statistically distinct patterns of retention and/or creation of the single-site mC to T mutants within the forward and reverse strand 24 nt LTR7 sequence in human genome (bottom panel). Panel **C** (top left panel) shows evolutionary conservation in primates of chromosomal positions of the 213 single site mC to T mutants of 24 nt LTR7 sncRNA-encoding sequences and numbers of various single-site mC to T mutants of the reverse strand 24 nt LTR7 sncRNA-encoding sequence in human genome (panel **C**, top right panel).

Panel **E** documents highly correlated patterns of chromosomal distributions in human genome of the 33 nt (top figure) and 24 nt (bottom figure) in length wild-type LTR7 sncRNA-encoding sequences (r = 0.771; p < 0.0001)

Panel **F** (top panel) demonstrates co-localization of 89% of unique to human NANOG-binding sites expressing LTR7 sncRNAs with DNA methylation sites in hESC.

Panel **F** (bottom panels) shows statistically distinct patterns of retention and/or creation in human genome of the single-site mC to T mutants of 13 to 24 nt in length (3,963 sequences; bottom left figure) and 12 to 24 nt in length (14,066 sequences; bottom right figure) within the reverse strand 24 nt LTR7 sncRNA-encoding sequence.

**Supplemental Figure S5.** Related to Figures 1 & 3.

Multiple sequence alignments of unique to human NANOG-binding sites embedded within LTR and LINE sequences identifies motif logos and positions of nucleotides with high frequency conservation within NANOG-binding LTR7-derived sequences expressed in hESC.

Motif logos and positions of nucleotides with high frequency conservation within unique to human NANOG-binding sites embedded within LTR and LINE sequences (**A**) and within NANOG-binding LTR7-derived sequences expressed in hESC (**B, C**). Note similarity of motif logos of unique to human NANOG-binding sites and previously reported (9, 42) consensus POU5F1 (OCT4) and NANOG-binding sites (**A**). Large box in **B** (bottom right panel) shows the zoom-in view of the 33 nt LTR7 sncRNA motif logos. Underlined nucleotides in **B** (left panel) depict positions of sequences resembling motif logos of TF-binding sites associated with NANOG-binding LTR7-derived sequences expressed in hESC.

Figure **C** shows zoom-in view of motif logos associated with NANOG-binding LTR7-derived sequences expressed in hESC. Overall heights of the stacks of symbols at each position indicate the sequence conservation at that position; the height of each symbol within the stack indicates the relative frequency of the particular nucleotide at that position. Arrows indicate positions of the putative NANOG-binding sequences within 200 nt sequences of 165 LTR-embedded unique to human NANOG-binding sites (**A**, top left panel). Consensus TF-binding sites motif logos were previously reported (9, 28, 32).

**Supplemental Figure S6.** Related to Figure 4.

24 brain-specific and 53 housekeeping rapidly-evolving protein-coding genes analyzed in this study manifest increased evolutionary rates in primates (**A**) and enriched proximity placements of unique to human NANOG-binding sites (**B, G, H**).

Mirror-image reciprocity between the numbers of unique to human NANOG- and CTCF-binding events near the pluripotency and lineage specification lncRNAs (**C**) reflecting distinct patterns of placement and/or retention of unique to human TF-binding sites near fifteen pluripotency lncRNAs compared to fourteen lineage specification lncRNAs (**D**). Panels **E** and **F** show the correlation plots reflecting distinct patterns of association of rapidly evolving in primates brain-specific and housekeeping genes with unique to human NANOG-binding sites: the placement of unique to human NANOG-binding sites is enriched near brain-specific genes having highest Ka/Ks ratios defining most rapidly evolving genes (**E**) and housekeeping genes with the lowest Ka/Ks ratios reflecting their lowest rates of protein evolution (**F**).

**Supplemental Figure S7.** Related to Figure 5.

hESC genomic maps of NANOG-binding sites, adjacent 5hmC, and functional hESC enhancer/gene pairs involved in enhancer-coding gene interactions.

Snapshots of base-resolution 5hmC maps in H1-hESC genome in close proximity to the unique to human NANOG-binding sites located near functional hESC enhancers involved in high-confidence enhancer-target gene interactions reported by Xie et al. (2012). Only predicted enhancer-gene pairs with reported p-value <=0.0001 were considered in the analysis (Xie et al., 2012). Also shown are genomic positions of the unique to human NANOG-binding sites (horizontal red bars), hESC enhancers, transcription start sites (TSS) of enhancer-targeted protein-coding genes, and maps of hESC enhancer-gene interaction pairs in H1-hESC (semi-transparent green arks). Positive blue bar values indicate 5hmC contents on the (+) Watson strand of DNA, whereas negative values indicate 5hmC contents on the (-) Crick strand. For 5hmC values, the vertical axis limits are -50% to +50%. Note that unique to human NANOG-binding sites (red horizontal bars marked by the NANOG signs) are placed in the heart of the large-scale hESC regulatory domains near functional distal enhancers within high-complexity multi-dimensional regulatory networks of interacting enhancer-gene pairs (**A**) as well as within bi-directional (**B**) distal regulatory domains. Examples of 5hmC patterns identified near unique to human NANOG-binding sites on all human chromosomes are shown in **C** and **D**. The exact genomic positions of 5hmC are indicated by blue vertical bars, NANOG-binding sites coordinates are depicted by red horizontal bars (unique to human loci), red arrows (primate-specific sequences), and red stars (common with rodents sites) based on hg18 release of the human genome reference database. Note that unique to human and primate-specific NANOG-binding sequences are located near hESC enhancers, whereas common with rodents NANOG-binding sites are placed next to TSS of protein-coding genes. Examples of common 5hmC three to eight letter symbols utilized to create a large variety of 5hmC patterns near unique to human NANOG-binding sites are shown in the yellow box (**C**, top left panel).

**Supplemental Figure S8.** Related to Figure 5.

Classification of 5hmC patterns near unique to human NANOG-binding sites shown in Figs. 5 and S7.

**Supplemental Figure S9.** Related to Figures 1-5.

Genomic and transcriptional landscapes near unique to human CTCF, OCT4, NANOG binding sites embedded within transcriptionally active in human cells retrotransposons.

Example 1: Human Feb. 2009 (GRCh37/hg19) chrX:136,520,161-136,520,840 (680 bp).

**Supplemental Figure S10.** Related to Figures 1-5.

Genomic and transcriptional landscapes near unique to human CTCF, OCT4, NANOG binding sites embedded within transcriptionally active in human cells retrotransposons.

Example 2: Human Feb. 2009 (GRCh37/hg19) chr5:147,248,497-147,249,099 (603 bp).

**Supplemental Figure S11.** Related to Figures 1-5.

Genomic and transcriptional landscapes near unique to human CTCF, OCT4, NANOG binding sites embedded within transcriptionally active in human cells retrotransposons.

Example 3: Human Feb. 2009 (GRCh37/hg19) chrX:93,959,399-93,959,698 (300 bp).

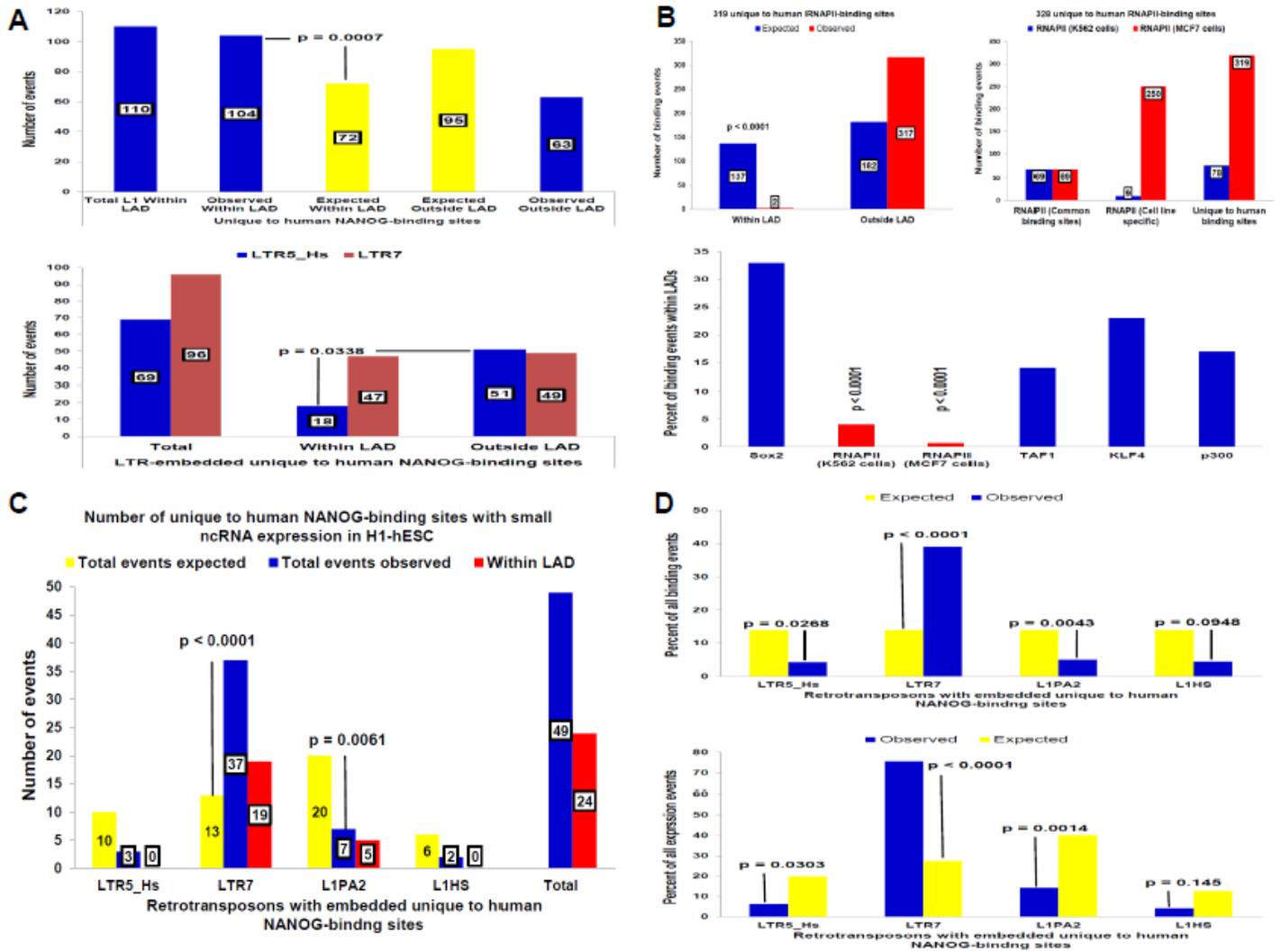

**Figure 1.**

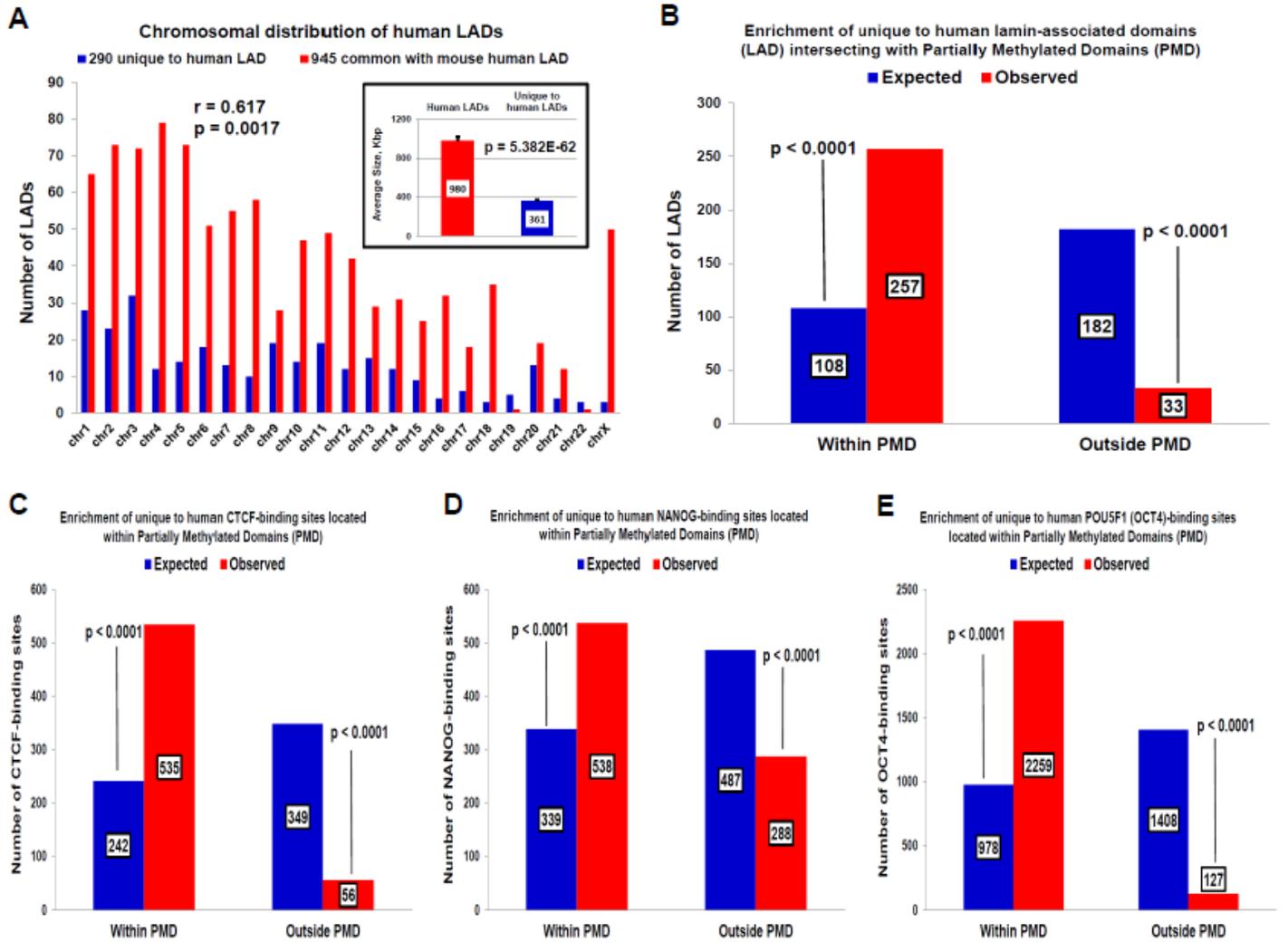

**Figure 2.**

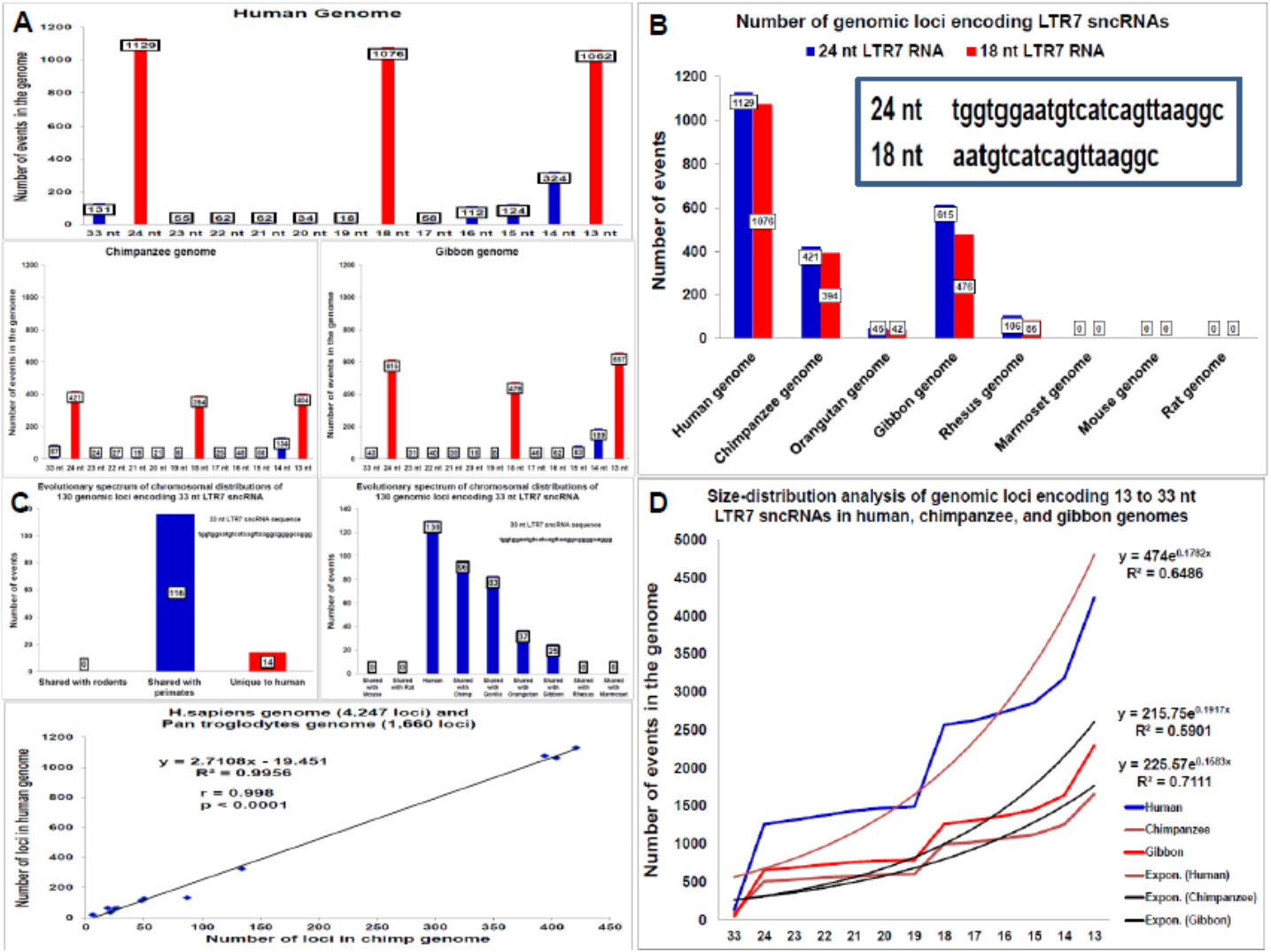

**Figure 3.**

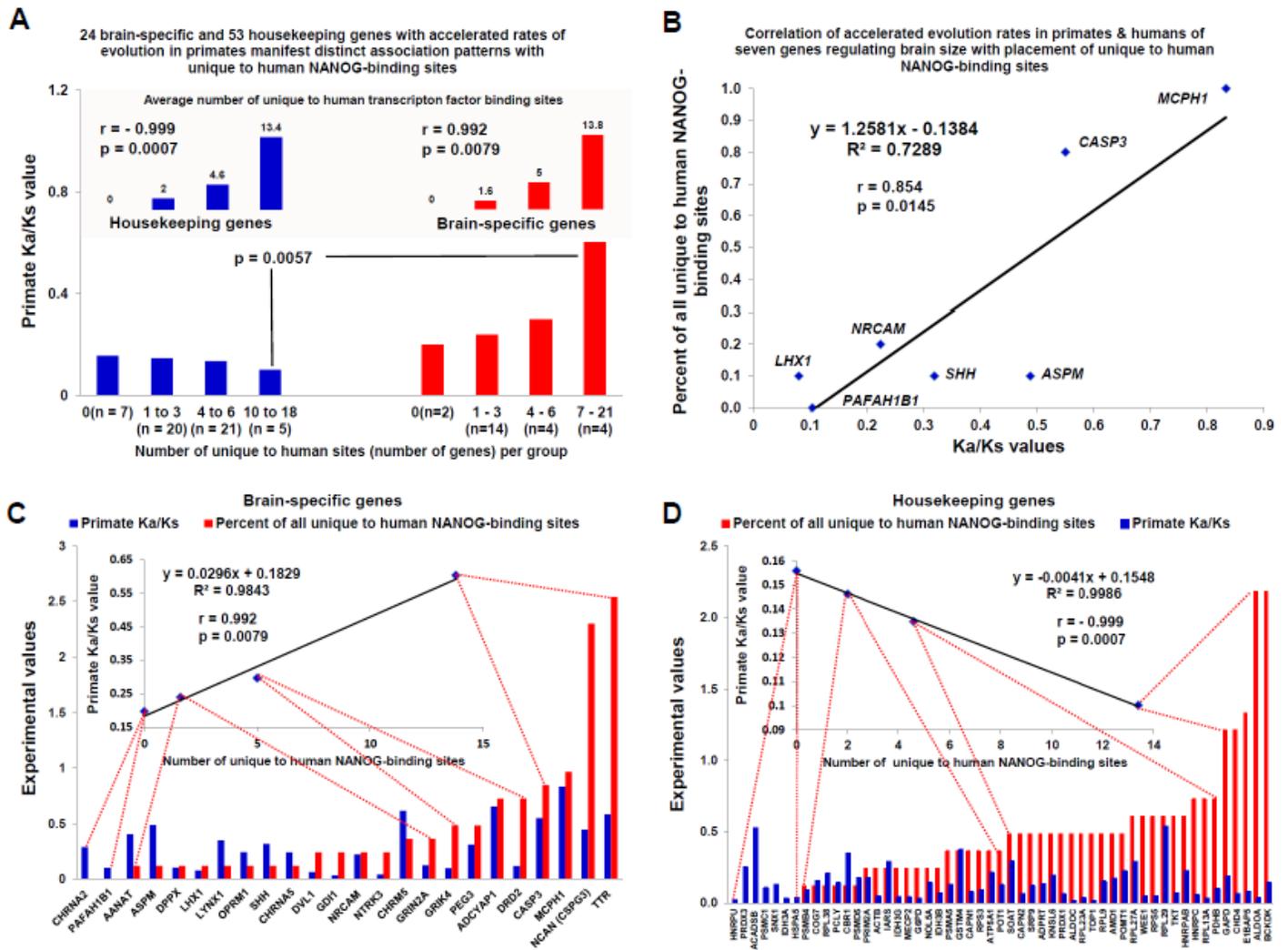

**Figure 4.**

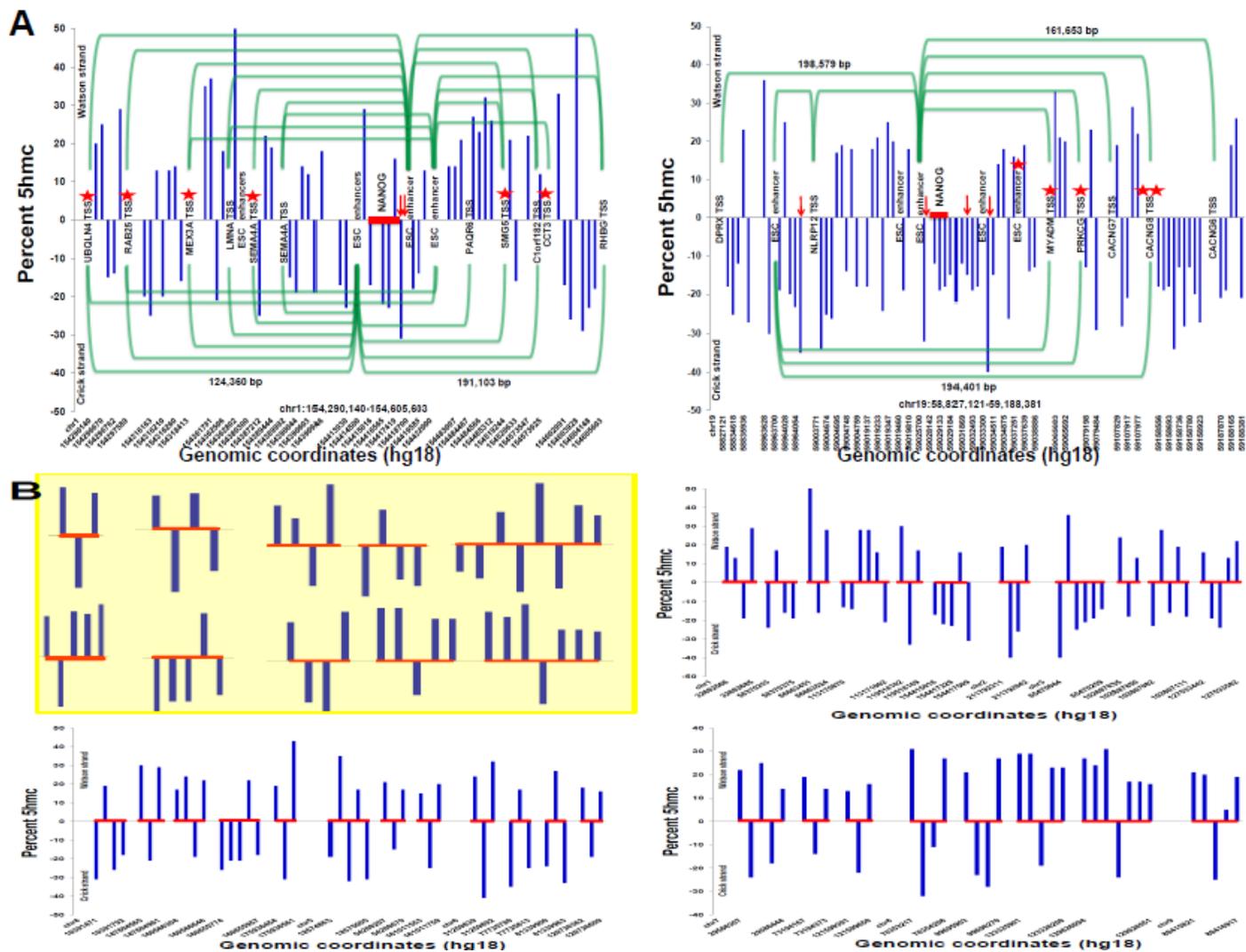

**Figure 5.**